\documentclass{pasj01}
\usepackage[usenames,dvipsnames]{xcolor}

\newcommand\nii{[N~{\sc ii}]~}
\newcommand\oiii{[O~{\sc iii}]~}

\begin{document} 
\Received{}
\Accepted{}

\title{Environmental impacts on molecular gas in protocluster galaxies at $z\sim2$}

\author{Ken-ichi \textsc{Tadaki}\altaffilmark{1}}%
\altaffiltext{1}{National Astronomical Observatory of Japan, 2-21-1 Osawa, Mitaka, Tokyo 181-8588, Japan}
\email{tadaki.ken@nao.ac.jp}
\author{Tadayuki \textsc{Kodama}\altaffilmark{2}}
\author{Masao \textsc{Hayashi}\altaffilmark{1}}
\author{Rhythm \textsc{Shimakawa}\altaffilmark{3}}
\author{Yusei \textsc{Koyama}\altaffilmark{3,4}}
\author{Minju \textsc{Lee}\altaffilmark{5}}
\author{Ichi \textsc{Tanaka}\altaffilmark{3}}
\author{Bunyo \textsc{Hatsukade}\altaffilmark{6}}
\author{Daisuke \textsc{Iono}\altaffilmark{1,4}}
\author{Kotaro \textsc{Kohno}\altaffilmark{6,7}}
\author{Yuichi \textsc{Matsuda}\altaffilmark{1,4}}
\author{Tomoko \textsc{Suzuki}\altaffilmark{1,2}}
\author{Yoichi \textsc{Tamura}\altaffilmark{5}}
\author{Jun \textsc{Toshikawa}\altaffilmark{8}}
\author{Hideki \textsc{Umehata}\altaffilmark{9}}
\altaffiltext{2}{Astronomical Institute, Tohoku University, Aoba-ku, Sendai 980-8578, Japan}
\altaffiltext{3}{Subaru Telescope, National Astronomical Observatory of Japan, National Institutes of Natural Sciences, 650 North A'ohoku Place, Hilo, HI 96720}
\altaffiltext{4}{Department of Astronomical Science, SOKENDAI (The Graduate University for Advanced Studies), Mitaka, Tokyo 181-8588, Japan}
\altaffiltext{5}{Division of Particle and Astrophysical Science, Nagoya University, Furocho, Chikusa, Nagoya 464-8602, Japan}
\altaffiltext{6}{Institute of Astronomy, Graduate School of Science, The University of Tokyo, 2-21-1 Osawa, Mitaka, Tokyo 181-0015, Japan}
\altaffiltext{7}{Research Center for the Early Universe, Graduate School of Science, The University of Tokyo, 7-3-1 Hongo, Bunkyo-ku, Tokyo 113-0033, Japan}
\altaffiltext{8}{Institute for Cosmic Ray Research, The University of Tokyo, Kashiwa, Chiba 277-8582, Japan.}
\altaffiltext{9}{The Institute of Physical and Chemical Research (RIKEN), 2-1 Hirosawa, Wako-shi, Saitama 351-0198, Japan}




\maketitle

\begin{abstract}
We present the results from ALMA CO(3--2) observations of 66 H$\alpha$-selected galaxies in three protoclusters around radio galaxies, PKS1138-262 ($z=2.16$) and USS1558-003 ($z=2.53$), and 4C23.56 ($z=2.49$).
The pointing areas have an overdensity of $\sim$100 compared to a mean surface number density of galaxies in field environments.
We detect CO emission line in 16 star-forming galaxies, including previously published six galaxies, to measure the molecular gas mass.
In the stellar mass range of $10.5<\log(M_\mathrm{star}/M$\solar$)<11.0$, the protocluster galaxies have larger gas mass fractions and longer gas depletion timescales compared to the scaling relations established by field galaxies.
On the other hand, the amounts of molecular gas in more massive galaxies with $\log(M_\mathrm{star}/M$\solar$)>11.0$ are comparable in mass to the scaling relation, or smaller.
Our results suggest that the environmental effects on gas properties are mass-dependent: 
in high-density environments, gas accretion through cosmic filaments is accelerated in less massive galaxies 
while this is suppressed in the most massive system.
\end{abstract}

\section{Introduction}

In the local Universe at $z<0.1$, most galaxies in massive clusters occupy the red sequence on the color-magnitude diagram, implying that they are quenched, passively evolving galaxies (e.g., \cite{1984ApJ...285..426B}).
Massive, red sequence galaxies likely first emerge in protoclusters at $z=2-3$ when the cosmic star formation history peaks \citep{2007MNRAS.377.1717K}.
Therefore, protoclusters in the forming phase are an appropriate laboratory for studying the origin to form the massive-end of the red sequence.
CO measurements of protocluster galaxies are critical for assessing the molecular gas reservoirs fueling star formation.
Efficient gas accretion in less massive halos retains or enhances the ability to form stars (e.g., \cite{2009Natur.457..451D}) while ram-pressure stripping in clusters removes hot gas, fueling star formation, from galaxies (e.g., \cite{2016A&A...591A..51S}).
On the other hand, field galaxies are subject to influence by time-variations of gas accretion.
A key question is whether the environmental variations in molecular gas properties exceed the time-variations seen in field galaxies.

For field galaxies, \citet{2018ApJ...853..179T} establish the scaling relations of gas mass fraction, $f_\mathrm{gas}=M_\mathrm{gas}/(M_\mathrm{star}+M_\mathrm{gas})$, and gas depletion timescale, $\tau_\mathrm{depl}=M_\mathrm{gas}$/SFR, by compiling CO data for 667 galaxies over $0 < z < 4$, predicting CO luminosities from redshift, stellar mass and star formation rates (SFR) relative to the main sequence at fixed stellar mass, SFR/SFR$_\mathrm{MS}$ (see also \cite{2015ApJ...800...20G}).
In clusters or protocluster regions, recent observations have detected the CO emission from a few dozen galaxies at $z=1.5-2.5$ \citep{2012MNRAS.426..258A, 2017ApJ...841L..21H, 2018ApJ...856..118H, 2017ApJ...842L..21N, 2017ApJ...849..154S, 2017ApJ...849...27R, 2014ApJ...788L..23T, 2016Sci...354.1128E, 2018ApJ...867L..29W, 2017ApJ...842...55L, 2017A&A...608A..48D,2018MNRAS.479..703C}.
However, many of these previous studies focus on only one cluster.

Here we present the results from CO $J=3-2$ observations of three protoclusters, PKS 1138-262 (PKS1138) at $z=2.16$, USS 1558-003 (USS1558) at $z=2.53$ and 4C 23.56 (4C23) at $z=2.49$, using Atacama Large Millimeter/submillimeter Array (ALMA).
We assume a Chabrier initial mass function (IMF; \cite{2003PASP..115..763C}) and adopt cosmological parameters of $H_0$ =70 km s$^{-1}$ Mpc$^{-1}$, $\Omega_{\rm M}$=0.3, and $\Omega_\Lambda$ =0.7.

\section{Sample selection}

\subsection{H$\alpha$ emitter survey with Subaru}

Three protoclusters of PKS1138, USS1558 and 4C23 were originally identified as an excess of Ly$\alpha$ emitters and/or distant red galaxies around the radio galaxy \citep{2000A&A...358L...1K, 2006MNRAS.371..577K, 2007MNRAS.377.1717K, 1997ApJS..109..367K}.
In these protoclusters, we have made a systematic H$\alpha$ emitter survey with narrow-band filters (\cite{2013IAUS..295...74K}), using MOIRCS near-infrared instrument on Subaru.
Narrow-band filters could pick up \oiii emitters at $z\sim3$ as well as H$\alpha$ emitters at $z\sim2$.
Utilizing the difference in the Balmer break wavelengths in the observed-frame, we have efficiently removed the interlopers based on the color-color magnitudes \citep{2016ApJ...826L..28H,2018MNRAS.481.5630S}.
Our panoramic H$\alpha$ mapping identifies an X-shaped structure centered on the radio galaxy in PKS1138 \citep{2013MNRAS.428.1551K,2018MNRAS.481.5630S}, two dense star-bursting groups of galaxies in USS1558 \citep{2012ApJ...757...15H,2016ApJ...826L..28H,2018MNRAS.473.1977S} and a 2 Mpc-scale group on the east side of the radio galaxy in 4C23 \citep{2011PASJ...63S.415T}.
These structures are also spectroscopically confirmed to be associated with the radio galaxies \citep{2014MNRAS.441L...1S,2011PASJ...63S.415T}.
In this work, we do not use 10 H$\alpha$ emitters with X-ray detection \citep{2002A&A...396..109P,2011MNRAS.412..705B,2018arXiv180506569M} because the SFR estimate is highly uncertain due to Active Galactic Nuclei (AGN) contribution to H$\alpha$ luminosity.

\begin{figure*}[t]
\begin{center}
\includegraphics[width=160mm]{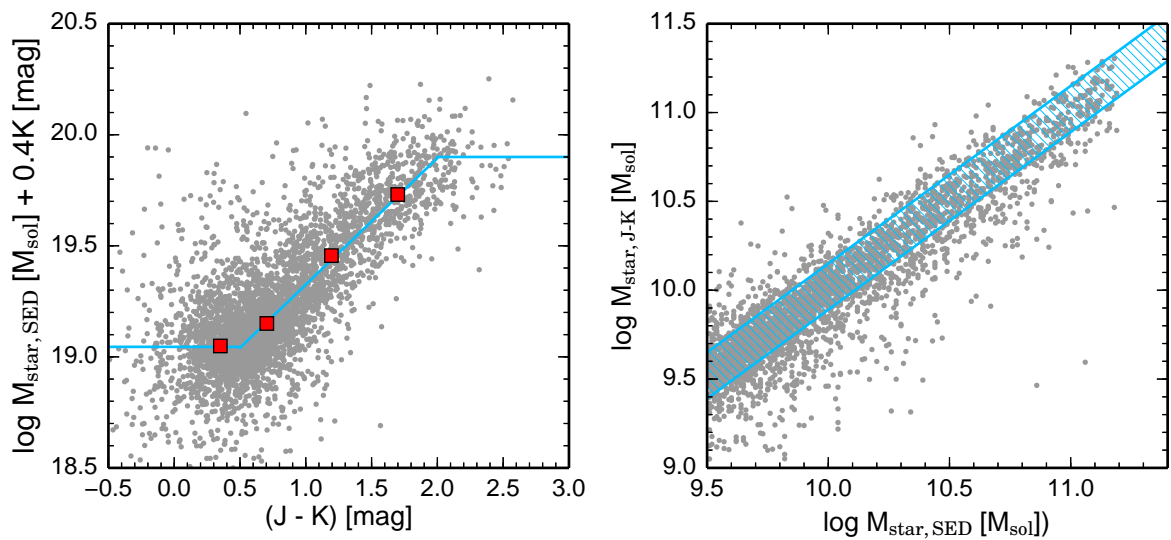}
\end{center}
\caption{
Left: stellar mass-to-light ratios as a function of $J-K$ color for $\sim$3000 star-forming galaxies at $z=2.1-2.6$ in field environments \citep{2014ApJS..214...24S, 2016ApJS..225...27M}.
Red squares show the median values in each color bin and a cyan line indicates the best-fit relation given by the equation (1).
Right: a comparison between SED fitting-based and single color-based stellar masses. A cyan hatched region shows the standard deviation ($\pm0.13$ dex) of the differences ($\log M_\mathrm{star,SED}/\log M_\mathrm{star,J-K}$).
}
\label{fig;JK}
\end{figure*}

\subsection{Stellar mass and star formation rate}
\label{sec;mstar_sfr}

For our sample of star-forming galaxies in three protoclusters, we estimate stellar mass and star formation rate (SFR) in the same manner.
Spectral energy distribution (SED) fitting would give better estimates of stellar mass while the results depend on the number of the used photometric band and the depths.
We adopt a more simple method to minimize the difference in richness of the available data among three protoclusters.
At $z\sim2.5$, the $J-K$ color is a good estimate of stellar mass-to-light ratio as the Balmer break is shifted to between the two bands.
Using SED fitting-based stellar masses, $K$-band magnitudes, and $J-K$ colors for $\sim$3000 star-forming galaxies at $z=2.1-2.6$ in the 3D-HST catalog \citep{2014ApJS..214...24S, 2016ApJS..225...27M}, 
we derive a simple relation by linear fitting (Figure \ref{fig;JK}) as,

\begin{equation}
\log (M_\mathrm{star,SED}/M$\solar$)+ 0.4 K= 0.57\times(J-K)+18.76.
\end{equation}

\noindent
We set $J-K=0.5$ and 2.0 to the minimum and maximum value as the mass-to-light ratios plateau in these ranges.
In the stellar mass range of $9.5<\log (M_\mathrm{star}/M$\solar$)<11.5$, the median and the standard deviation of differences between single color-based and SED fitting-based stellar masses are 0.02 dex and 0.13 dex, respectively (Figure \ref{fig;JK}).
We incorporate this dispersion in errors of stellar mass measurements.
We note that there remain some uncertainties in the stellar mass estimates if protocluster galaxies do not follow the same relation as field galaxies.

We use a {\it Spitzer}/MIPS 24 $\mu$m image to identify dusty star formation and estimate total infrared (IR) luminosities of HAEs. 
The data is retrieved from the Spitzer Heritage Archive (AORKEYs are 14888704, 14888960, 14889216 for 1138, 21852416 for 1558, and 10765568, 17910528 for 4C23). 
We apply the self-calibration on the BCD images to remove large-scale sky patterns and make mosaic images with a pixel scale of 1.25 arcsec, using MOPEX software \citep{2005ASPC..347...81M}. 
Source extraction and 7\arcsec aperture photometry are performed with APEX module in MOPEX.
We use an aperture correction of 2.57 to derive total fluxes, following the recipe in the MIPS Data Handbook.
The 3$\sigma$ limiting fluxes are 43 $\mu$Jy in 1138,  86 $\mu$Jy in 1558 and 50 $\mu$Jy in 4C23, which correspond to SFR=46 $M_\odot$yr$^{-1}$, 174 $M_\odot$yr$^{-1}$, and 92 $M_\odot$yr$^{-1}$, respectively.

Next, we extrapolate total infrared luminosities, $L_\mathrm{IR}$, with MIPS 24 $\mu$m fluxes and estimate SFRs \citep{1998ARA&A..36..189K}.
24 $\mu$m is not an ideal wavelength to accurately derive total infrared luminosities. 
\citet{2011A&A...533A.119E} shows comparisons between MIPS 24 $\mu$m-based and Herschel-based infrared luminosities using \citet{2001ApJ...556..562C} templates. 
In the range of $\log(L_\mathrm{IR}/L_\odot)>12$, the 24 $\mu$m method overestimates infrared luminosities. 
\citet{2011ApJ...738..106W} find that 24 $\mu$m-based infrared luminosities are well correlated with Herschel-based ones if \citep{2008ApJ...682..985W} templates are used.
Therefore, we use \citet{2008ApJ...682..985W} templates to avoid the overestimate of SFRs.

For galaxies without detection at mid-infrared, SFRs are computed from H$\alpha$ luminosities with correction for dust extinction.
We correct for contribution of \nii line to narrow-band fluxes assuming a mass-metallicity relation of star-forming galaxies at similar redshift \citep{2014ApJ...795..165S}.
A combination of stellar mass and H$\alpha$ equivalent width predicts the dust extinction level ($A_{\mathrm{H}\alpha}$) with the systematic uncertainty of 0.32 mag \citep{2015MNRAS.453..879K}.
This prediction is calibrated by using local galaxies, but \citet{2018MNRAS.476.3218C} find that the dust properties like a dust extinction-stellar mass relation do not evolve significantly from local galaxies out to high-redshift.
We propagate the measurement errors of observed H$\alpha$ luminosities, $L(\mathrm{H}\alpha)_\mathrm{uncor}$, and the systematic uncertainty of dust extinction, $A_{\mathrm{H}\alpha}$, to the errors in the estimate of intrinsic H$\alpha$ luminosities, $L(\mathrm{H}\alpha)_\mathrm{cor}$, on the basis of the equation of $L(\mathrm{H}\alpha)_\mathrm{cor}$=$L(\mathrm{H}\alpha)_\mathrm{uncor}\times10^{(0.4A_{\mathrm{H}\alpha})}$.

\begin{figure*}[t]
\begin{center}
\includegraphics[width=160mm]{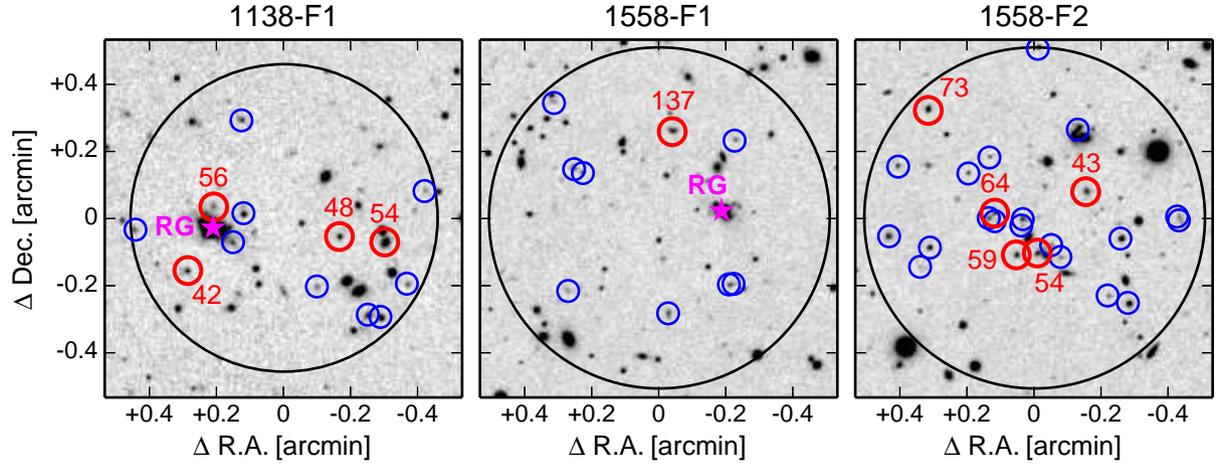}
\end{center}
\caption{2D spatial distributions of H$\alpha$ emitters in protoclusters at $z\sim2$ on Subaru/MOIRCS narrow-band images.
Red and blue circles represent CO-detected and non-detected objects, respectively.
Only for the CO-detected galaxies, their IDs are noted near the circles.
Magenta stars indicate the positions of radio galaxies.
The primary beam of the ALMA pointing is shown by large black circles.
}
\label{fig;FoV}
\end{figure*}

\begin{table*}[t]
\caption{Properties of 16 H$\alpha$-selected star-forming galaxies with CO(3--2) line detections. \label{sourcelist}}
\begin{center}
\begin{tabular}{lcccccccccc}
\hline
 ID & R.A. & Decl. & $\log~M_\mathrm{star}$ & SFR & $z_\mathrm{CO}$ &  FWHM & $S_\mathrm{CO}dV$  & $\log M_\mathrm{gas}$ & $f_\mathrm{gas}$ & $\tau_\mathrm{depl}$\\
 & (deg) & (deg) &  ($M$\solar$$) & ($M$\solar$$yr$^{-1}$) & & (km s$^{-1}$) &  (Jy km s$^{-1}$) & ($M$\solar$$) & & (Gyr) \\
\hline
1138.42 & 175.2029 & -26.4879 & 9.79 $\pm$ 0.15 & 41 $\pm$ 12 & 2.163 & 185 & 0.19 $\pm$ 0.04 & 11.21 & 0.96 & 4.0 \\
1138.48 & 175.1945 & -26.4862 & 10.65 $\pm$ 0.13 & 144 $\pm$ 66 & 2.157 & 232 & 0.42 $\pm$ 0.03 & 10.99 & 0.69 & 0.7 \\
1138.54 & 175.1919 & -26.4865 & 11.18 $\pm$ 0.13 & 466 $\pm$ 214 & 2.148 & 328 & 0.88 $\pm$ 0.04 & 11.23 & 0.53 & 0.4 \\
1138.56 & 175.2014 & -26.4847 & 9.90 $\pm$ 0.15 & 36 $\pm$ 11 & 2.144 & 224 & 0.20 $\pm$ 0.04 & 11.08 & 0.94 & 3.4 \\
1558.43 & 240.2942 & -0.5189 & 11.08 $\pm$ 0.13 & 66 $\pm$ 21 & 2.528 & 698 & 0.26 $\pm$ 0.04 & 10.84 & 0.37 & 1.1 \\
1558.54 & 240.2966 & -0.5219 & 10.35 $\pm$ 0.13 & 67 $\pm$ 20 & 2.515 & 242 & 0.23 $\pm$ 0.02 & 11.00 & 0.82 & 1.5 \\
1558.59 & 240.2977 & -0.5220 & 10.72 $\pm$ 0.13 & 88 $\pm$ 26 & 2.513 & 420 & 0.49 $\pm$ 0.02 & 11.18 & 0.75 & 1.7 \\
1558.64 & 240.2988 & -0.5199 & 9.83 $\pm$ 0.14 & 50 $\pm$ 15 & 2.529 & 721 & 0.21 $\pm$ 0.04 & 11.54 & 0.98 & 6.9 \\
1558.73 & 240.3021 & -0.5149 & 10.33 $\pm$ 0.13 & 106 $\pm$ 31 & 2.526 & 261 & 0.19 $\pm$ 0.03 & 10.91 & 0.79 & 0.8 \\
1558.137 & 240.3247 & -0.4756 & 10.51 $\pm$ 0.13 & 98 $\pm$ 29 & 2.525 & 264 & 0.16 $\pm$ 0.02 & 10.76 & 0.64 & 0.6 \\
4C23.3 & 316.8376 & 23.5206 & 10.86 $\pm$ 0.13 & 182 $\pm$ 84 & 2.488 & 482 & 0.45 $\pm$ 0.08 & 11.11 & 0.64 & 0.7 \\
4C23.4 & 316.8401 & 23.5281 & 11.15 $\pm$ 0.13 & 379 $\pm$ 174 & 2.490 & 281 & 0.33 $\pm$ 0.05 & 10.93 & 0.38 & 0.2 \\
4C23.8 & 316.8164 & 23.5243 & 10.74 $\pm$ 0.13 & 166 $\pm$ 76 & 2.486 & 197 & 0.50 $\pm$ 0.05 & 11.17 & 0.73 & 0.9 \\
4C23.9 & 316.8441 & 23.5287 & 10.67 $\pm$ 0.14 & 318 $\pm$ 146 & 2.485 & 739 & 0.77 $\pm$ 0.11 & 11.38 & 0.84 & 0.8 \\
4C23.10 & 316.8154 & 23.5200 & 10.55 $\pm$ 0.15 & 99 $\pm$ 32 & 2.485 & 303 & 0.47 $\pm$ 0.09 & 11.20 & 0.82 & 1.6 \\
4C23.16 & 316.8110 & 23.5211 & 10.54 $\pm$ 0.17 & 205 $\pm$ 94 & 2.484 & 544 & 0.70 $\pm$ 0.12 & 11.38 & 0.87 & 1.2 \\
\hline
\end{tabular}
\end{center}
\end{table*}

\section{Observations and results}

\subsection{Target fields}

We have carried out new ALMA observations in PKS1138 and USS1558 protoclusters with three pointings including the two radio galaxies (1138-F1 and 1558-F1) and the densest group in the southwest of USS1558 (1558-F2).
Figure \ref{fig;FoV} shows the spatial distributions of a total of 46 H$\alpha$ emitters in the observed regions.
We define the environment of these protoclusters, based on the 5th-nearest-neighbour local surface number density of H$\alpha$ emitters ($\Sigma_\mathrm{5th}$).
To determine the mean density of galaxies in field environments ($\Sigma_\mathrm{5th,mean}$), we use a large sample of H$\alpha$ emitters from High-Z Emission Line Survey (HiZELS; \cite{2013MNRAS.428.1128S}).
We adopt the same criteria of H$\alpha$ equivalent width and luminosity both in MAHALO and HiZELS sample to minimize the impact of the sample selection (EW$_\mathrm{rest}>40$ \AA~ and $\log (L_{\mathrm{H}\alpha}/\mathrm{ergs~s}^{-1})>42.2$).
Then, the local over-density is computed as excess to the mean density, 

\begin{equation}
\delta = \frac{\Sigma_\mathrm{5th} - f_\mathrm{cor}\Sigma_\mathrm{5th, mean}}{f_\mathrm{cor}\Sigma_\mathrm{5th, mean}}.
\end{equation}

\noindent
We also correct for the difference of the widths of narrow-band filters between the MAHALO and the HiZELS ($f_\mathrm{cor}=1.24$).
Both surveys trace H$\alpha$ emitters at $z\sim2$, but the HiZELS filter probes a wider survey volume in the redshift direction.
The observed regions in the PKS1138 and USS1558 protoclusters typically have overdensities of  $\delta =121$ and $145$, respectively.
Such high-density regions are not seen in the HiZELS survey area of 2 deg$^2$.

\begin{figure*}[!h]
\begin{center}
\includegraphics[width=160mm]{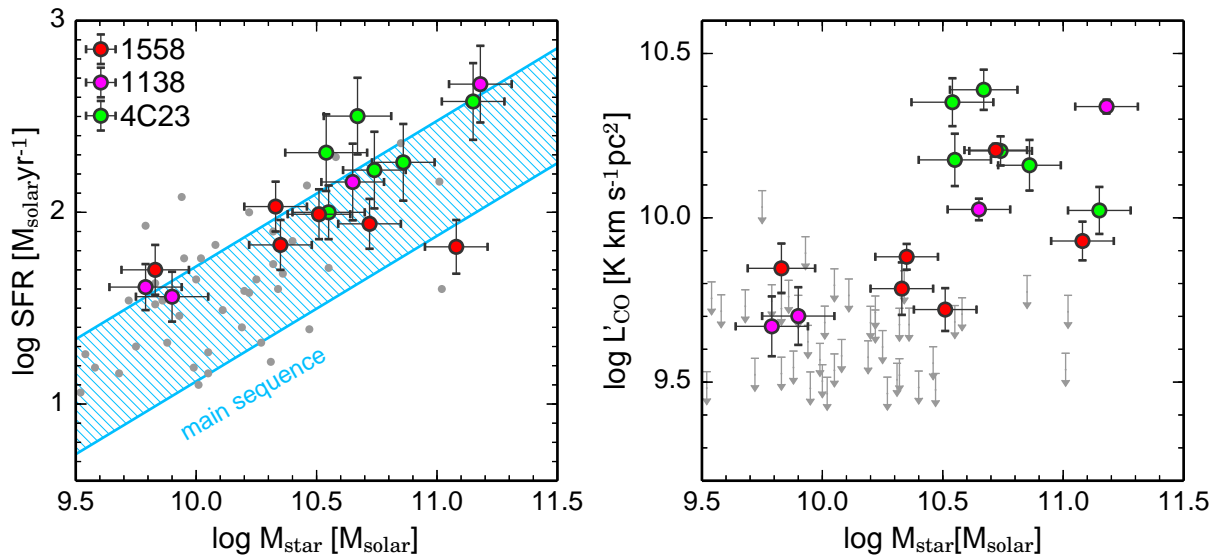}
\end{center}
\caption{Left: stellar mass vs. SFR for CO sample in protoclusters at $z\sim2$ (red circles in USS1558, magenta circles in PKS1138, and green circles in 4C23.56). A cyan shaded region shows the range of the main sequence of star-forming galaxies with a 1$\sigma$ scatter of 0.3 dex \citep{2014ApJS..214...15S}.
Gray dots show CO non-detected H$\alpha$ emitters.
Right: stellar mass vs. CO line luminosity. 
The upper limit is determined by the 5$\sigma$ values of CO fluxes assuming a velocity width of 400 km s$^{-1}$.
}
\label{fig;MS}
\end{figure*}

\subsection{ALMA observations}

To observe the CO (3--2) emission line ($\nu_\mathrm{rest}=345.796$ GHz in the rest-frame) at $z\sim2.5$, we use the Band-3 receivers with the 64-input correlator in Time Division Mode in a central frequency of 110 and 98 GHz.
On-source time is 1.3 hours in PKS1138 and 2.4 hours per pointing in USS1558.
We utilize the Common Astronomy Software Application package ({\tt CASA}; \cite{2007ASPC..376..127M}) for the data calibration.
We reconstruct clean maps with channel width of 100 km s$^{-1}$, adopting natural weighting.
In the 1138-F1, 1558-F1, and 1558-F2 cubes, the synthesized beamsizes are \timeform{1.8"}$\times$\timeform{1.5"}, \timeform{1.8"}$\times$\timeform{1.4"}, and \timeform{2.0"}$\times$\timeform{1.5"}, respectively.
The rms levels are 125 $\mu$Jy beam$^{-1}$, 90 $\mu$Jy beam$^{-1}$, and 97 $\mu$Jy beam$^{-1}$ per 100 km s$^{-1}$ bin.

For 4C23 protocluster, \citet{2017ApJ...842...55L} have made CO(3--2) observations to cover 21 H$\alpha$ emitters with four ALMA pointings.
The median overdensity is $\delta =79$.
We make clean maps in the same way as PKS1138 and USS1558 data.
With natural weighting, the synthesized beamsize and the rms level are \timeform{0.9"}$\times$\timeform{0.7"} and 174 $\mu$Jy beam$^{-1}$ per 100 km s$^{-1}$ bin, respectively.
We also smooth the maps to match the beam size to \timeform{1.5"} for measurements of galaxy-integrated fluxes.

\subsection{CO detections}

\begin{figure*}[!h]
\begin{center}
\includegraphics[width=160mm]{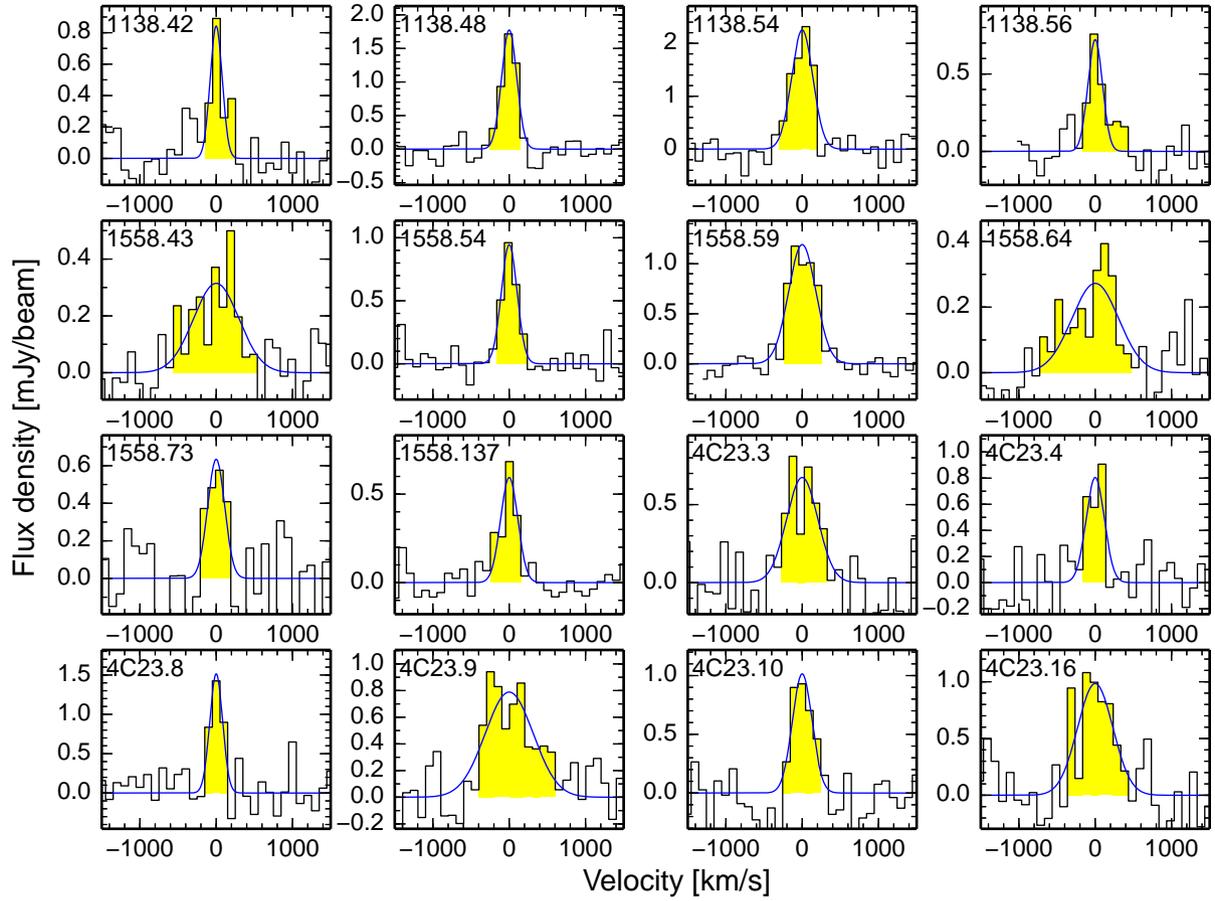}
\end{center}
\caption{The CO spectra of H$\alpha$-selected star-forming galaxies in three protoclsuters. The blue lines show the best-fitting profile with single Gaussian model. The integrated velocity ranges are shown by yellow shaded regions
}
\label{fig;spectra}
\end{figure*}

We use velocity-binned cubes with a width of 400 km s$^{-1}$ to search for a $>$5$\sigma$ peak within a 2\arcsec aperture around the H$\alpha$ emitters in the frequency range expected from the narrow-band redshift ($\Delta z=\pm0.02$).
We also have searched for a negative 5$\sigma$ peak in the same region, but did not detect any negative signals, indicating that our criterion is safe. 
We have detected the CO line from a total of 16 galaxies (4 in PKS1138, 6 in USS1558 and 6 in 4C23).
\citet{2017ApJ...842...55L} report CO detection in 4C23.5, but this object does not satisfy the detection criterion adopted in this work.
For four galaxies (1138.54, 1558.43, 1558.54, 1558.59), the CO (1--0) emission line was previously detected in ATCA \citep{2017A&A...608A..48D} or JVLA observations \citep{2016Sci...354.1128E, 2014ApJ...788L..23T}.
All CO(3--2)-detected galaxies in PKS1138 were also detected in CO(4--3) \citep{2018MNRAS.477L..60E}.
Seven galaxies (1138.48, 1138.54, 4C23.3, 4C23.4, 4C23.8, 4C23.9, 4C23.16) are detected at MIPS 24 $\mu$m.
For the remaining 9 galaxies, their H$\alpha$-based SFRs are consistent within the error with the upper limit given by the MIPS 24 $\mu$m data (Section \ref{sec;mstar_sfr}).

The CO detection rate is high, 69\% (11/16), in the stellar mass range of $\log(M_\mathrm{star}/M$\solar$)>10.5$ while it decreases to 12\% (6/50) for lower stellar mass galaxies (Figure \ref{fig;MS}). 
The low detection rate is probably due to a combination of small gas masses and metallicity effects on CO-to-H$_2$ conversion factor \citep{2013ARA&A..51..207B, 2012ApJ...746...69G}.
In this stellar mass range, our CO sample is likely to be biased to active star-forming galaxies in the upper side of the main sequence (Figure \ref{fig;MS}).

Figure \ref{fig;spectra} presents the CO spectra extracted from the peak position in the clean cubes.
Fitting the CO (3--2) spectra with single Gaussian function, we find that the line width is over a wide range of FWHM=180--740 km s$^{-1}$ (Table \ref{sourcelist}). 
We made moment-0 maps in the velocity range where the flux density is above 15 \% of the peak of the best-fit Gaussian spectra to measure the velocity-integrated CO fluxes.
A flux integrated in an inadequate range of velocity would be boosted or de-boosted by noise fluctuations \citep{2018MNRAS.479..703C}.
To investigate this effect, we create the simulated spectra by adding the best-fit Gaussian models to 35 noise spectra extracted in sky regions in 1558-F2 field as a representative.
We then measure the velocity-integrated fluxes in the same way as mentioned above and derive the ratios of the model flux to the measured flux.
The median values range from 0.97 to 1.03, indicating little systematic boosting effects.

\section{Molecular gas properties}
\label{sec;gasprop}

\begin{figure*}[!h]
\begin{center}
\includegraphics[width=160mm]{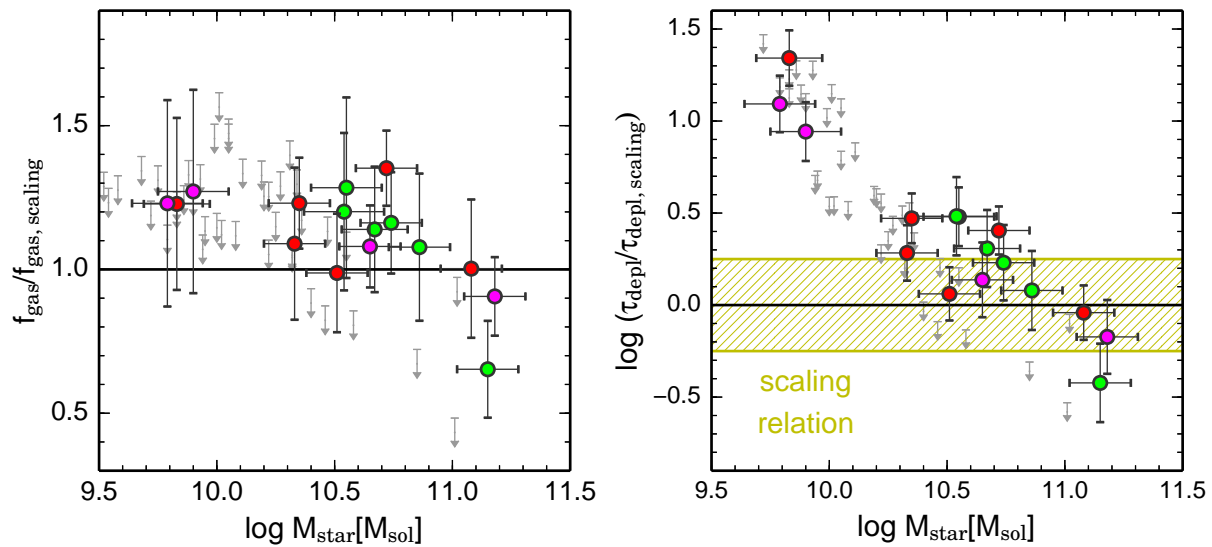}
\end{center}
\caption{
The ratio of gas mass fraction and gas depletion timescale to the predicted values from the scaling relations, as a function of stellar mass.
The uncertainty of the scaling relation is shown by the yellow shaded region ($\pm0.25$ dex).
The error bars take into account the uncertainties of stellar mass, SFR and gas mass estimates.
}
\label{fig;fgas}
\end{figure*}

We estimate the gas masses in the protocluster galaxies by following the method by \citet{2018ApJ...853..179T}.
Molecular gas masses are usually estimated from CO($J$=1--0) line luminosities, $L'_{(1-0)}$, while our ALMA observations provide the CO($J$=3--2) line luminosities, $L'_{(3-2)}$.
The CO line luminosity ratio approaches unity when gas density and/or temperature are sufficiently high.
The environmental dependence of CO excitation would be an interesting issue while probing it requires observations of lower-$J$ CO line.
Therefore, we assume the same CO excitation with $L'_{(1-0)}/L'_{(3-2)}$=1.8, which is a typical value in field galaxies \citep{2013ARA&A..51..207B, 2012ApJ...746...69G}, for a fair comparison.
We also adopt a metallicity-dependent CO-to-H$_2$ conversion factor of $M_\mathrm{gas}/L'_{(1-0)}=4.36\times f_\mathrm{metal}~(M$\solar (K km s$^{-1}$pc$^{-2}$)$^{-1}$) where $f_\mathrm{metal}$ corrects for metallicity effects through the stellar mass-metallicity relation \citep{2013ARA&A..51..207B, 2012ApJ...746...69G}.
The correction factor is $f_\mathrm{metal}=1.1$ at $\log(M_\mathrm{star}/M$\solar$)=11$ and $f_\mathrm{metal}=1.4$ at $\log(M_\mathrm{star}/M$\solar$)=10.5$.
We also give the 5$\sigma$ upper limit of molecular gas mass using CO cubes with 400 km s$^{-1}$ width for galaxies without CO detections (Table \ref{nonsourcelist} in the Appendix).

The gas mass fractions and the gas depletion timescales are listed in Table \ref{sourcelist}.
As gas mass fractions depend on stellar mass and SFR \citep{2018ApJ...853..179T, 2017ApJ...837..150S, 2015ApJ...800...20G}, we compare the measured molecular gas properties with the scaling relations at the fixed stellar mass and SFR.
The scaling relation of \citet{2018ApJ...853..179T} is derived from 667 field galaxies, including starburst galaxies as well as normal star-forming galaxies.
Their SFR offsets relative to the main-sequence are distributed over the range of $-0.5 $ dex$<\log(\mathrm{SFR}/\mathrm{SFR}_\mathrm{MS})<+1.0$ dex.
For extreme starburst outliers or quiescent galaxies, the scaling relation is not verified yet.
Our sample of protocluster galaxies is within the valid range of the scaling relation.
The gas mass fractions are still larger and the gas depletion timescales are significantly longer compared to the scaling relations (Figure \ref{fig;fgas}).
This is partly due to sample bias and the sensitivity limit of our observations.
In the stellar mass range of $\log(M_\mathrm{star}/M$\solar$)<10.5$, only five out of 50 galaxies are detected in CO.
Therefore, they are particularly gas-rich outliers, resulting in enhancing the star formation activity with $\log$(SFR/SFR$_\mathrm{MS})\sim0.3$ dex.

In the stellar mass range of $10.5<\log(M_\mathrm{star}/M$\solar$)<11.0$, 
the excess of gas mass fractions cannot be explained neither by sample bias nor higher SFRs 
because the detection rate of CO is high, 73\% (=8/11), and the galaxies are mostly located on the main-sequence in the stellar mass-SFR diagram (Figure \ref{fig;MS}).
Nevertheless, the gas depletion timescales are longer than expected from the scaling relation by 0.5 dex. 
The protocluster galaxies tend to have larger gas reservoirs as seen in cluster galaxies at $z\sim1.5$ \citep{2017ApJ...842L..21N}.

In contrast, we do not find such a gas-rich system in the stellar mass range of $\log(M_\mathrm{star}/M$\solar$)>11.0$.
Three CO-detected galaxies (1138.54, 1558.43, 4C23.4) are moderately gas-rich with $f_\mathrm{gas}=0.37-0.53$ and their gas depletion timescales are consistent with the scaling relation (Figure \ref{fig;fgas}).
We also comment that one non-detected galaxy (1138.58) has an extremely short gas depletion timescale of $\tau_\mathrm{depl}<0.2$ Gyr due to active star formation.
This galaxy is likely in the transition phase from star-forming to quiescent.

\section{Discussion}
\label{sec;discussion}

At a fixed redshift and stellar mass, galaxies with higher SFR have a larger gas mass fraction, suggesting that the rate of gas accretion onto galaxies is higher as well \citep{2012ApJ...758...73S, 2017ApJS..233...22S}.
Efficient gas accretion increases the amount of available gas in galaxies, which results in enhancing the star formation activity.
The scaling relation takes into account time-variations of gas accretion for 667 galaxies with CO measurements but does not include galaxies in high-density environments.
Exploiting CO measurements for 16 star-forming galaxies associated to protoclusters at $z\sim2.5$, 
we derive the gas mass fractions and the gas depletion time scales.
The environmental difference in molecular gas properties is likely dependent of stellar mass of the system although we need a larger sample to verify the scaling relation in high-density environments.

In the stellar mass range of $10.5<\log(M_\mathrm{star}/M$\solar$)<11.0$, we find protocluster galaxies to have larger gas mass fractions than expected from the scaling relation.
Such an enhanced gas fraction is reported in massive clusters at $z\sim1.5$ \citep{2018ApJ...856..118H, 2017ApJ...842L..21N}.
In contrast, \citet{2018ApJ...860..111D} report no environmental dependence of gas properties for galaxies at $z\sim2$ in COSMOS field, but their sample does not include extremely high-density regions with $\delta\sim$100.
Our results in the proto-clusters suggest the existence of more efficient gas accretion in regions at the intersection of cosmic filaments \citep{2017MNRAS.468L..21S}.
Nevertheless, they still remain on the main-sequence of star-forming galaxies, resulting in longer gas depletion timescales.
The variation of gas depletion time scales would be related to the spatial distributions of molecular gas within galaxies.
As star formation occurs only in dense, cold molecular clouds, a supply of newly accreted gas does not necessarily enhance star formation.
In this work, we use CO(3--2) emission line as a tracer of a total amount of molecular gas including diffuse one, under assumption of a constant CO(3--2)/CO(1--0) ratio although it in principle probes dense ($n>10^4$ cm$^{-3}$) or warm ($T>$30 K) gas. 
Therefore, it is not clear how much gas is directly associated with star formation. 
If starburst preferentially occurs in central compact regions and molecular gas is more extended \citep{2017ApJ...841L..25T},
the galaxy-integrated gas depletion timescales could be longer compared to the case when the molecular gas is compact as well.
Higher-resolution CO observations would be needed to look at the origin of the environmental variations of gas depletion timescales.

On the other hand, we find no excess of gas mass fraction in the most massive system with $\log(M_\mathrm{star}/M$\solar$)>11.0$.
The deficiency of molecular gas requires a physical process to suppress efficient gas accretion only in the most massive system.
Numerical simulations show that there are two modes of gas accretion (cold with T$<10^{5}$ K and hot with T$\sim10^{6}$ K; \cite{2005MNRAS.363....2K,2006MNRAS.368....2D}).
In less massive halos with $\log(M_\mathrm{halo}/M$\solar$)<12$, dense cold gas is directly accreted onto galaxies through streams, leading to starbursts.
By the steady gas supply, high-redshift star-forming galaxies can maintain active star formation and stay on the main-sequence.
In massive halos with $\log(M_\mathrm{halo}/M$\solar$)>12$, the accreted gas is heated by a virial shock and the remaining gas in galaxies is consumed by star formation. 
The halo quenching would also affect gas properties of lower stellar mass galaxies in massive halos, 
but the current sensitivity is not high enough to identify the deficient CO emission.
Another possible mechanism is AGN feedback.
AGN-driven outflows preferentially expel low density gas, leading to decrease of gas fractions, while they have little effect on dense star-forming gas \citep{2014MNRAS.441.1615G}.
If AGN activities are increased in massive galaxies in high-density environments \citep{2012PASJ...64...22T}, 
the gas fractions would be smaller than those of field galaxies.

It is still challenging to test these scenarios and other possibilities such as ram pressure stripping (e.g., \cite{1989ApJ...344..171K,2014A&A...564A..67B}).
Our ALMA observations demonstrate great potential to succeed statistical CO studies in protocluster galaxies with $\log(M_\mathrm{star}/M$\solar$)>10.5$.
Given the high signal-to-noise ratios in some galaxies, higher-resolution observations allow us to spatially resolve the CO emission.
The dynamical properties like baryonic mass fraction and rotation velocity-to-dispersion ratio could shed light on the physical processes responsible for the environmental impacts on molecular gas properties.

\bigskip

\begin{ack}
We thank the referee for constructive comments.
This paper makes use of the following ALMA data: ADS/JAO.ALMA\#2015.1.00395.S. ALMA is a partnership of ESO (representing its member states), NSF (USA) and NINS (Japan), together with NRC (Canada) and NSC and ASIAA (Taiwan) and KASI (Republic of Korea), in cooperation with the Republic of Chile. The Joint ALMA Observatory is operated by ESO, AUI/NRAO and NAOJ.
The MIR data used in this paper are taken with the Spitzer Space Telescope, which is operated by the Jet Propulsion Laboratory, California Institute of Technology under a contract with NASA.
K.T. acknowledges support by Grant-in-Aid for JSPS Research Fellow JP17J04449 and Scientific Research (S) JP17H06130. 
Data analysis was in part carried out on the common-use data analysis computer system at the Astronomy Data Center (ADC) of the National Astronomical Observatory of Japan.
\end{ack}

\appendix

\section*{Catalog of CO non-detected galaxies in three protoclusters}
\label{sec;appendix}
For protocluster galaxies without CO detections, we give the 5$\sigma$ upper limit of molecular gas mass, gas mass fraction and gas depletion timescales, assuming a line width of 400 km s$^{-1}$ (table \ref{nonsourcelist}). Some galaxies do not have a meaningful upper limit on gas depletion timescale as the 5$\sigma$ value is more than 10 Gyr.

\begin{center}
\begin{longtable}{lcccccccc}
\caption{Properties of 52 H$\alpha$-selected star-forming galaxies without CO(3--2) line detections. \label{nonsourcelist}}
\hline
 ID & R.A. & Decl. & $\log~M_\mathrm{star}$ & SFR  & $S_\mathrm{CO}dV$  & $\log M_\mathrm{gas}$ & $f_\mathrm{gas}$ & $\tau_\mathrm{depl}$\\
 & (deg) & (deg) &  ($M$\solar$$) & ($M$\solar$$yr$^{-1}$) &  (Jy km s$^{-1}$) & ($M$\solar$$) & & (Gyr) \\
\hline
\endfirsthead
\hline
 ID & R.A. & Decl. & $\log~M_\mathrm{star}$ & SFR  & $S_\mathrm{CO}dV$  & $\log M_\mathrm{gas}$ & $f_\mathrm{gas}$ & $\tau_\mathrm{depl}$\\
 & (deg) & (deg) &  ($M$\solar$$) & ($M$\solar$$yr$^{-1}$) &  (Jy km s$^{-1}$) & ($M$\solar$$) & & (Gyr) \\
\hline
\endhead
\hline
\endfoot
\endlastfoot
1138.35 & 175.1922 & -26.4902 & 10.22 $\pm$ 0.13 & 101 $\pm$ 29 & $<$0.23 & $<$10.88 & $<$0.82 & $<$0.8 \\
1138.37 & 175.1957 & -26.4887 & 10.05 $\pm$ 0.15 & 14 $\pm$ 5 & $<$0.15 & $<$10.82 & $<$0.86 & $<$4.6 \\
1138.39 & 175.1929 & -26.4901 & 10.32 $\pm$ 0.13 & 79 $\pm$ 30 & $<$0.21 & $<$10.80 & $<$0.75 & $<$0.8 \\
1138.41 & 175.1907 & -26.4885 & 9.58 $\pm$ 0.17   & 16 $\pm$ 5 & $<$0.23 & $<$11.76 & $<$0.99 & - \\
1138.45 & 175.2004 & -26.4865 & 10.00 $\pm$ 0.14 & 45 $\pm$ 16 & $<$0.14 & $<$10.84 & $<$0.87 & $<$1.5 \\
1138.49 & 175.2058 & -26.4859 & 9.54 $\pm$ 0.17 & 18 $\pm$ 6 & $<$0.25 & $<$11.92 & $<$1.00 & - \\
1138.57 & 175.1897 & -26.4839 & 9.68 $\pm$ 0.18 & 15 $\pm$ 7 & $<$0.24 & $<$11.53 & $<$0.99 & - \\
1138.58 & 175.1998 & -26.4850 & 10.40 $\pm$ 0.13 & 71 $\pm$ 21 & $<$0.14 & $<$10.57 & $<$0.60 & $<$0.5 \\
1138.65 & 175.1999 & -26.4804 & 10.25 $\pm$ 0.14 & 44 $\pm$ 13 & $<$0.18 & $<$10.76 & $<$0.77 & $<$1.3 \\
1558.31 & 240.2897 & -0.5201 & 10.22 $\pm$ 0.19 & 38 $\pm$ 12 & $<$0.16 & $<$10.92 & $<$0.83 & $<$2.2 \\
1558.32 & 240.2896 & -0.5202 & 10.36 $\pm$ 0.14 & 47 $\pm$ 15 & $<$0.16 & $<$10.84 & $<$0.75 & $<$1.4 \\
1558.35 & 240.2921 & -0.5244 &   9.94 $\pm$ 0.13 & 121 $\pm$ 36 & $<$0.14 & $<$11.18 & $<$0.95 & $<$1.2 \\
1558.38 & 240.2925 & -0.5212 & 11.01 $\pm$ 0.13 & 144 $\pm$ 42 & $<$0.12 & $<$10.51 & $<$0.24 & $<$0.2 \\
1558.39 & 240.2931 & -0.5240 & 10.19 $\pm$ 0.17 & 25 $\pm$ 8 & $<$0.13 & $<$10.85 & $<$0.82 & $<$2.8 \\
1558.44 & 240.2946 & -0.5158 & 10.46 $\pm$ 0.13 & 139 $\pm$ 41 & $<$0.12 & $<$10.67 & $<$0.62 & $<$0.3 \\
1558.46 & 240.2955 & -0.5221 &   9.52 $\pm$ 0.15 & 12 $\pm$ 4 & $<$0.10 & $<$12.20 & $<$1.00 & - \\
1558.47 & 240.2959 & -0.5215 &   9.04 $\pm$ 0.21 & 13 $\pm$ 4 & $<$0.10 & $<$17.32 & $<$1.00 & - \\
1558.52 & 240.2967 & -0.5118 &   9.86 $\pm$ 0.14 & 35 $\pm$ 10 & $<$0.20 & $<$11.45 & $<$0.98 & $<$8.1 \\
1558.55 & 240.2974 & -0.5205 & 10.27 $\pm$ 0.23 & 21 $\pm$ 8 & $<$0.10 & $<$10.68 & $<$0.72 & $<$2.3 \\
1558.57 & 240.2974 & -0.5202 & 10.02 $\pm$ 0.14 & 57 $\pm$ 17 & $<$0.10 & $<$10.91 & $<$0.89 & $<$1.4 \\
1558.60 & 240.2990 & -0.5172 &   9.23 $\pm$ 0.17 & 10 $\pm$ 3 & $<$0.11 & $<$14.42 & $<$1.00 & - \\
1558.61 & 240.2988 & -0.5203 & 10.47 $\pm$ 0.22 & 24 $\pm$ 9 & $<$0.10 & $<$10.59 & $<$0.57 & $<$1.6 \\
1558.65 & 240.2990 & -0.5202 &   9.95 $\pm$ 0.14 & 58 $\pm$ 18 & $<$0.10 & $<$11.02 & $<$0.92 & $<$1.8 \\
1558.66 & 240.3001 & -0.5180 &   9.30 $\pm$ 0.15 & 15 $\pm$ 5 & $<$0.11 & $<$13.71 & $<$1.00 & - \\
1558.71 & 240.3020 & -0.5217 & 10.08 $\pm$ 0.13 & 68 $\pm$ 20 & $<$0.13 & $<$10.96 & $<$0.88 & $<$1.3 \\
1558.72 & 240.3024 & -0.5226 &   9.21 $\pm$ 0.18 & 6 $\pm$ 2 & $<$0.14 & $<$14.75 & $<$1.00 & - \\
1558.77 & 240.3036 & -0.5177 & 10.20 $\pm$ 0.14 & 39 $\pm$ 12 & $<$0.16 & $<$10.95 & $<$0.85 & $<$2.3 \\
1558.79 & 240.3040 & -0.5211 & 10.55 $\pm$ 0.13 & 51 $\pm$ 16 & $<$0.16 & $<$10.76 & $<$0.62 & $<$1.1 \\
1558.130 & 240.3217 & -0.4832 & 8.79 $\pm$ 0.22 & 4 $\pm$ 1 & $<$0.11 & $<$25.51 & $<$1.00 & - \\
1558.131 & 240.3217 & -0.4760 & 9.88 $\pm$ 0.14 & 21 $\pm$ 7 & $<$0.12 & $<$11.20 & $<$0.95 & $<$7.6 \\
1558.132 & 240.3219 & -0.4832 & 9.72 $\pm$ 0.15 & 35 $\pm$ 11 & $<$0.11 & $<$11.52 & $<$0.99 & $<$9.5 \\
1558.134 & 240.3249 & -0.4846 & 10.32 $\pm$ 0.13 & 54 $\pm$ 16 & $<$0.11 & $<$10.71 & $<$0.71 & $<$0.9 \\
1558.145 & 240.3292 & -0.4777 & 10.31 $\pm$ 0.13 & 17 $\pm$ 6 & $<$0.11 & $<$10.70 & $<$0.71 & $<$3.0 \\
1558.147 & 240.3299 & -0.4835   & 9.99 $\pm$ 0.15 & 15 $\pm$ 5 & $<$0.13 & $<$11.05 & $<$0.92 & $<$7.3 \\
1558.148 & 240.3296 & -0.4775   & 9.83 $\pm$ 0.14 & 33 $\pm$ 10 & $<$0.11 & $<$11.27 & $<$0.97 & $<$5.6 \\
1558.151 & 240.3306 & -0.4742 & 10.01 $\pm$ 0.13 & 13 $\pm$ 5 & $<$0.16 & $<$11.14 & $<$0.93 & - \\
4C23.2 & 316.8407 & 23.5304 & 10.85 $\pm$ 0.13 & 229 $\pm$ 106 & $<$0.21 & $<$10.77 & $<$0.45 & $<$0.3 \\
4C23.5 & 316.8207 & 23.5085 & 10.58 $\pm$ 0.13 & 194 $\pm$ 89 & $<$0.18 & $<$10.77 & $<$0.61 & $<$0.3 \\
4C23.6 & 316.8395 & 23.5221 & 10.34 $\pm$ 0.14 & 40 $\pm$ 13 & $<$0.22 & $<$10.95 & $<$0.81 & $<$2.2 \\
4C23.7 & 316.8147 & 23.5271 & 11.02 $\pm$ 0.14 & 40 $\pm$ 13 & $<$0.18 & $<$10.68 & $<$0.32 & $<$1.2 \\
4C23.12 & 316.8122 & 23.5299 & 9.17 $\pm$ 0.19 & 227 $\pm$ 68 & $<$0.23 & $<$15.24 & $<$1.00 & - \\
4C23.13 & 316.8409 & 23.5283 & 9.79 $\pm$ 0.16 & 84 $\pm$ 26 & $<$0.20 & $<$11.54 & $<$0.98 & $<$4.1 \\
4C23.14 & 316.8324 & 23.5142 & 9.83 $\pm$ 0.16 & 43 $\pm$ 14 & $<$0.19 & $<$11.43 & $<$0.98 & $<$6.4 \\
4C23.15 & 316.8331 & 23.5190 & 9.90 $\pm$ 0.17 & 37 $\pm$ 13 & $<$0.18 & $<$11.30 & $<$0.96 & $<$5.5 \\
4C23.17 & 316.8234 & 23.5307 & 9.75 $\pm$ 0.16 & 20 $\pm$ 7 & $<$0.38 & $<$11.91 & $<$0.99 & - \\
4C23.18 & 316.8401 & 23.5337 & 9.93 $\pm$ 0.17 & 29 $\pm$ 10 & $<$0.27 & $<$11.43 & $<$0.97 & $<$9.4 \\
4C23.19 & 316.8425 & 23.5294 & 10.11 $\pm$ 0.18 & 31 $\pm$ 11 & $<$0.20 & $<$11.09 & $<$0.91 & $<$4.0 \\
4C23.20 & 316.8123 & 23.5224 & 10.05 $\pm$ 0.18 & 19 $\pm$ 7 & $<$0.22 & $<$11.18 & $<$0.93 & $<$8.1 \\
4C23.22 & 316.8244 & 23.5291 & 9.10 $\pm$ 0.31 & 25 $\pm$ 9 & $<$0.38 & $<$16.49 & $<$1.00 & - \\
4C23.23 & 316.8115 & 23.5218 & 9.27 $\pm$ 0.37 & 23 $\pm$ 9 & $<$0.24 & $<$14.15 & $<$1.00 & - \\
\hline
\end{longtable}
\end{center}


\bibliographystyle{apj}

\end{document}